\documentclass[aps,pra,amsmath,amssymb,11pt,final,tightenlines,twoside,onecolumn,nofootinbib,superscriptaddress,showkeys,showkeywords]{revtex4-2}

\usepackage[T2A]{fontenc}
\usepackage[utf8x]{inputenc}
\usepackage[russian,english]{babel}
\usepackage{graphicx}% Include figure files

\begin{document}

\selectlanguage{english}

\noindent {\it ASTRONOMY REPORTS, 2025, Volume , Issue }
\bigskip\bigskip  \hrule\smallskip\hrule
\vspace{35mm}


\title{SEARCH FOR THE POSSIBLE MEMBERS OF THE OPEN CLUSTER NGC 3532 WITH POOR ASTROMETRIC SOLUTIONS OF GAIA DR3}

\author{\bf\copyright~2025 \firstname{D.~I.}~\surname{Tagaev}}
\email{dima.tagaev@list.ru}
\affiliation{Ural Federal University, Ekaterinburg, Russia}

\author{\bf\firstname{A.~F.}~\surname{Seleznev}}
\email{anton.seleznev@urfu.ru}
\affiliation{Ural Federal University, Ekaterinburg, Russia}

\begin{abstract}
\vspace{3mm}
\received{12.06.24}
\revised{12.06.24}
\accepted{12.06.24} 
\vspace{3mm}

We performed star counts in the region of the open cluster NGC 3532.
The ranges of trigonometric parallaxes and proper motions containing all the stars of the cluster were determined using the stars with 5- and 6-parameter Gaia DR3 solutions.
The estimated radius of the cluster was $R_c = 178 \pm 3$ arcminutes and the number of cluster stars was $N_c = 2200 \pm 40$.
We estimate the number of stars with poor astrometric solutions that may be members of the cluster.
For this purpose, we analyze the surface density distribution of stars with two-parameter Gaia DR3 solutions, stars with the parameter RUWE$>$1.4, and stars with large relative errors of trigonometric parallaxes in the vicinity of the cluster.
We are looking for stars that fall within the area of the color-magnitude diagram occupied by probable members of the NGC 3532 cluster from the Hant\&Reffert sample.
The radial surface density profile plotted with such stars shows the concentration of stars toward the cluster center.
An analysis of the profile yields an estimate of 2150$\pm$230 stars that may be cluster members.
Thus, nearly one half of cluster members can be lost when the probable members are selected only by exact astrometric data of Gaia DR3.
Among these lost stars, there may be a significant number of unresolved binary and multiple systems.

\end{abstract}

\maketitle

\section{INTRODUCTION}

The successful work of the Gaia space mission \cite{Gaia} and its catalogs, the latest of which is Gaia DR3 \cite{GaiaDR3}, have absolutely changed the study of the Galaxy and its subsystems.
Significant successes have been achieved in the study of open star clusters \cite{Cantat-Gaudin2024}.
Large-scale stellar structures were discovered in the regions of modern and recent star formation, linking known open clusters \cite{Jerabkova2019,Beccari2020}.
The nature of the Alpha Centauri stellar stream was clarified \cite{Nikiforova2020}.
Extended coronae \cite{Meingast2021} and tidal tails \cite{Yeh2019,Roser2019,Nikiforova2020,Ye+2021,Bhattacharya2022} were discovered in a number of open clusters.
Progress was made in studying the dynamics of open clusters \cite{Danilov2020,Danilov2021a,Danilov2021b,Pang+2021}, in studying the population of binary and multiple stars in clusters \cite{Danilov2020,Jadhav+2021,Pang+2023}.

However, the most important point is the possibility of reliable identification of the probable members of open clusters based on trigonometric parallaxes and proper motions of stars from the Gaia catalogs.
Cantat-Gaudin with co-authors \cite{Cantat-Gaudin+2020} listed the probable members of a large number of clusters based on the Gaia DR2 catalog \cite{GaiaDR2}.
Lists of probable members of more than 7000 open clusters, cluster candidates and stellar groups based on the Gaia DR3 catalog are presented in \cite{H&R2023}, a detailed analysis of these results is given in \cite{H&R2024}.
Such samples are the basis for studying the kinematics and dynamics of clusters.

Nevertheless, some tasks, such as determination of the luminosity function and the mass function, require the complete samples of cluster members.
Meanwhile, the samples of the probable cluster members from \cite{Cantat-Gaudin+2020,H&R2023} and from other works are not complete.
Obviously, the limiting magnitude of stars cuts any sample.
This fact is common for any study.
The main problem arises due to selection of the probable cluster members by parallaxes and proper motions.
It is known that about 19\% of all stars in Gaia DR3 \cite{GaiaDR3} have only 2-parameter solutions (Gaia DR3 do not provide parallaxes and proper motions for these stars).
Some of these stars may be cluster members.
Further, when one selects probable cluster members at the base of Gaia DR3, he/she restricts the sample by the RUWE (renormalized unit weight error) parameter.
The limit RUWE<1.4 is usually used.
In addition, investigators often introduce a restraint on the relative parallax error(see, for example, \cite{El-Depsey+2023}).
The purpose of these restrictions is clear.
The authors want to obtain a sample of reliably selected cluster members.
But, in this case, the completeness of the sample inevitably suffers.

We can explain the reason for the appearance of stars with 2-parameter solutions, stars with negative parallax values, stars with large errors in astrometric parameters, stars with large RUWE values in Gaia DR3 by the fact that the astrometric parameters in Gaia DR3 are determined based on the model of the motion of a single star across the sky \cite{Lindegren+2021}.
The motion of binary and multiple stars with a separation between the components near the Gaia resolution limit of 0.2--1.5 arcseconds (Gaia Early Data Release 3, Documentation release 1.1; https://gea.esac.esa.int/archive/documentation/GEDR3/index.html) is much more complicated.
Therefore, the application of the single star motion model leads to large errors in the astrometric parameters.
Thus, a large number of unresolved binary and multiple stars could be among the cluster stars that had been lost during a compiling the samples of probable cluster members.

The aim of this paper is to estimate how many stars with "poor" astrometric solutions may be cluster members.
With this goal, we use the open cluster NGC 3532 as an example.
In Section II, we perform star counts in the NGC 3532 wide region based on stars with 5- and 6-parameter solutions of Gaia DR3 in order to estimate the radius and number of cluster stars.
In Section III, we look for stars with "poor" astrometric solutions that may be cluster members based on their position on the color-magnitude diagram (CMD).
Section IV is devoted to a discussion of the results of the paper.

\section{ESTIMATION OF THE RADIUS AND A STAR NUMBER OF NGC 3532 BY STARS WITH 5- AND 6-PARAMETRIC SOLUTIONS}

In this paper, we restrict our study to the limiting magnitude $G=18$ mag.
The reason is that for fainter stars the errors of the astrometric parameters in Gaia DR3 increase sharply.
Thus, at $G=18$ mag the lower value of the parallax error becomes greater than 0.1 mas, at $G=19$ mag — greater than 0.2 mas, at $G=20$ mag — greater than 0.4 mas.

\begin{figure}[ht]
\includegraphics[width=0.5\textwidth]{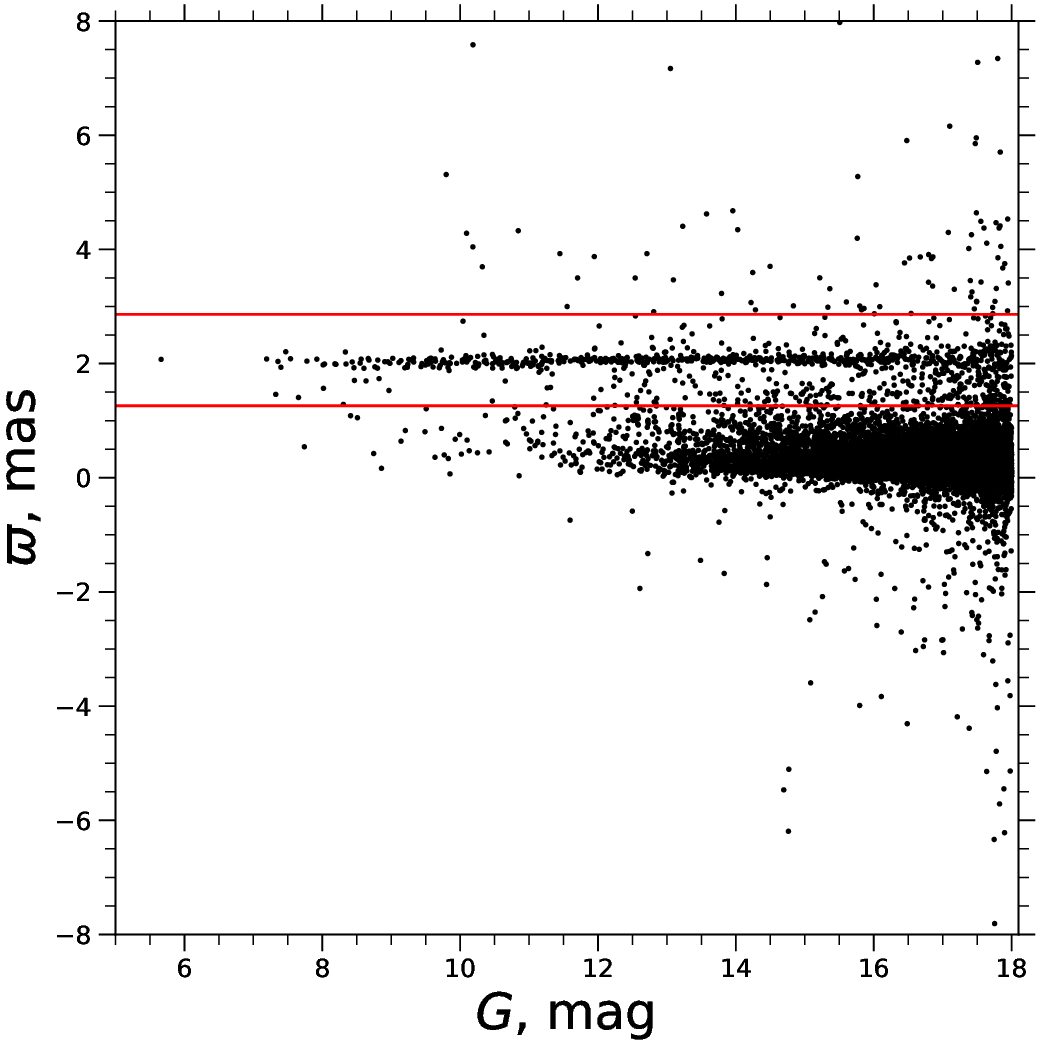}
\caption{Diagram `parallax $\varpi$ -- stellar magnitude G' for stars from the region of 0.5$^\circ \times$0.5$^\circ$ around the center of the cluster NGC 3532. The red lines show the parallax range in which the cluster stars are contained.}
\label{plx-G_1}
\end{figure}

\begin{figure}[ht]
\includegraphics[width=0.5\textwidth]{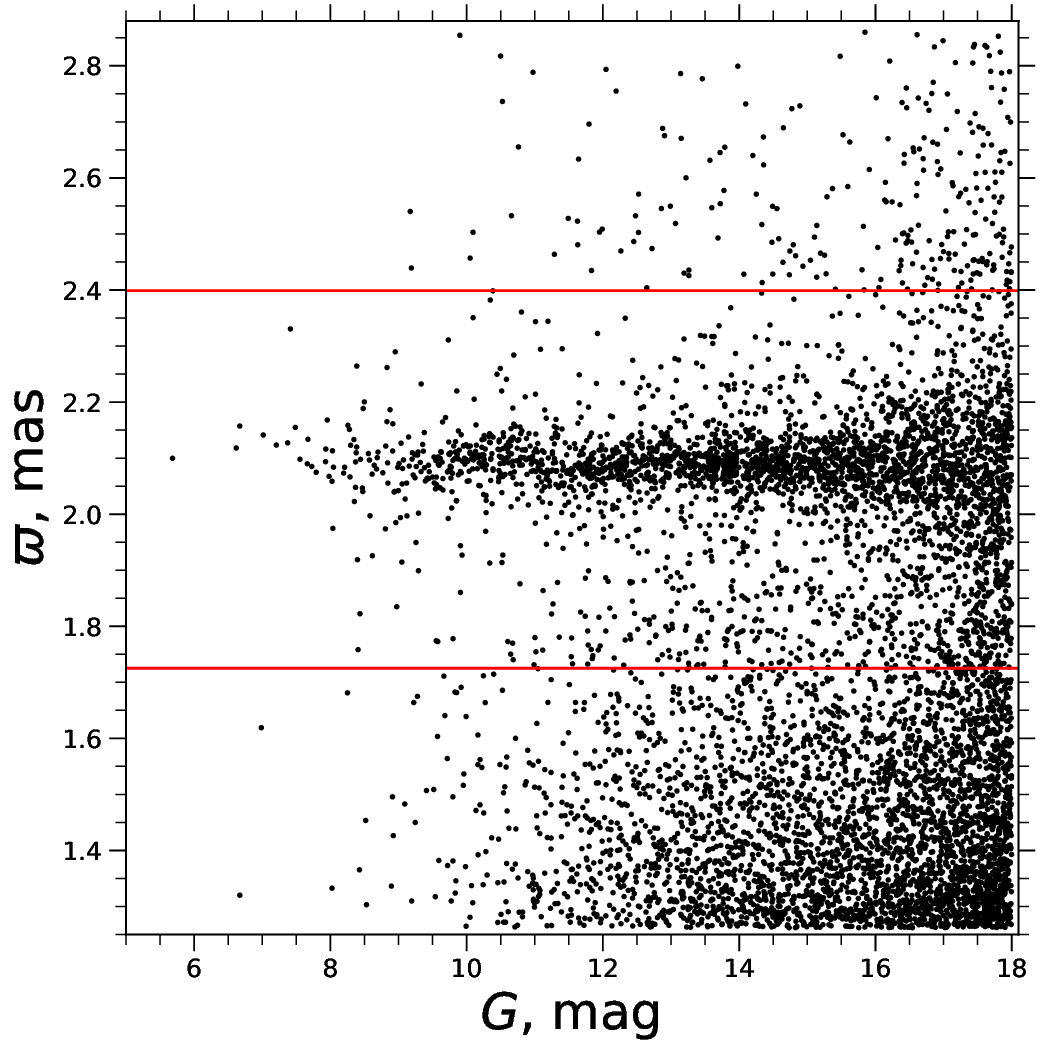}
\caption{Diagram `parallax $\varpi$ -- stellar magnitude G' for stars in the 6.0$^\circ \times$6.0$^\circ$ region around the center of the NGC 3532 cluster, satisfying the constraints on parallaxes and proper motions (\ref{interval_1}). The red lines show the refined parallax range.}
\label{plx-G_2}
\end{figure}

\begin{figure}[ht]
\includegraphics[width=0.7\textwidth]{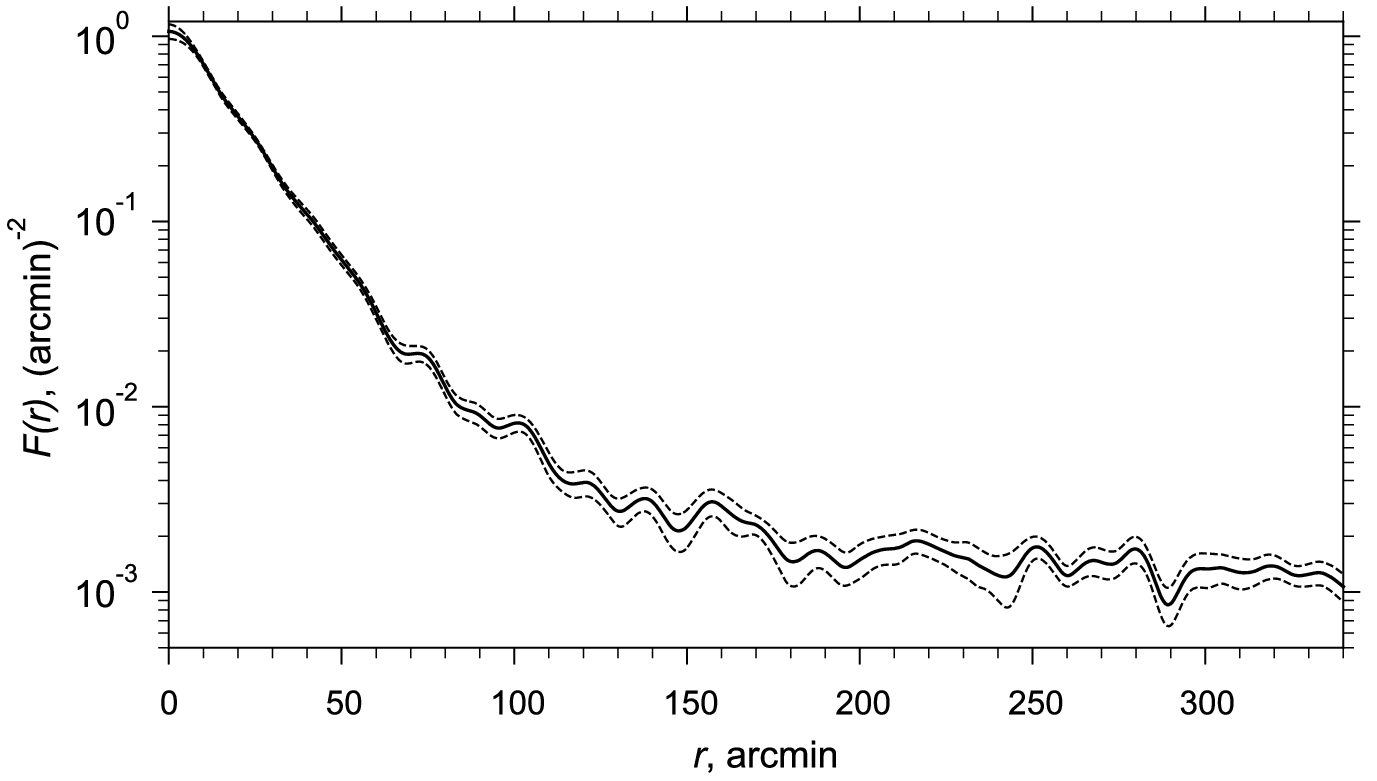}
\caption{Radial density profile of the NGC 3532 cluster plotted by the KDE with a kernel half-width of 10 arcmin. The dashed line shows the confidence interval of $2\sigma$-width. }
\label{profile}
\end{figure}

\begin{figure}[ht]
\includegraphics[width=0.7\textwidth]{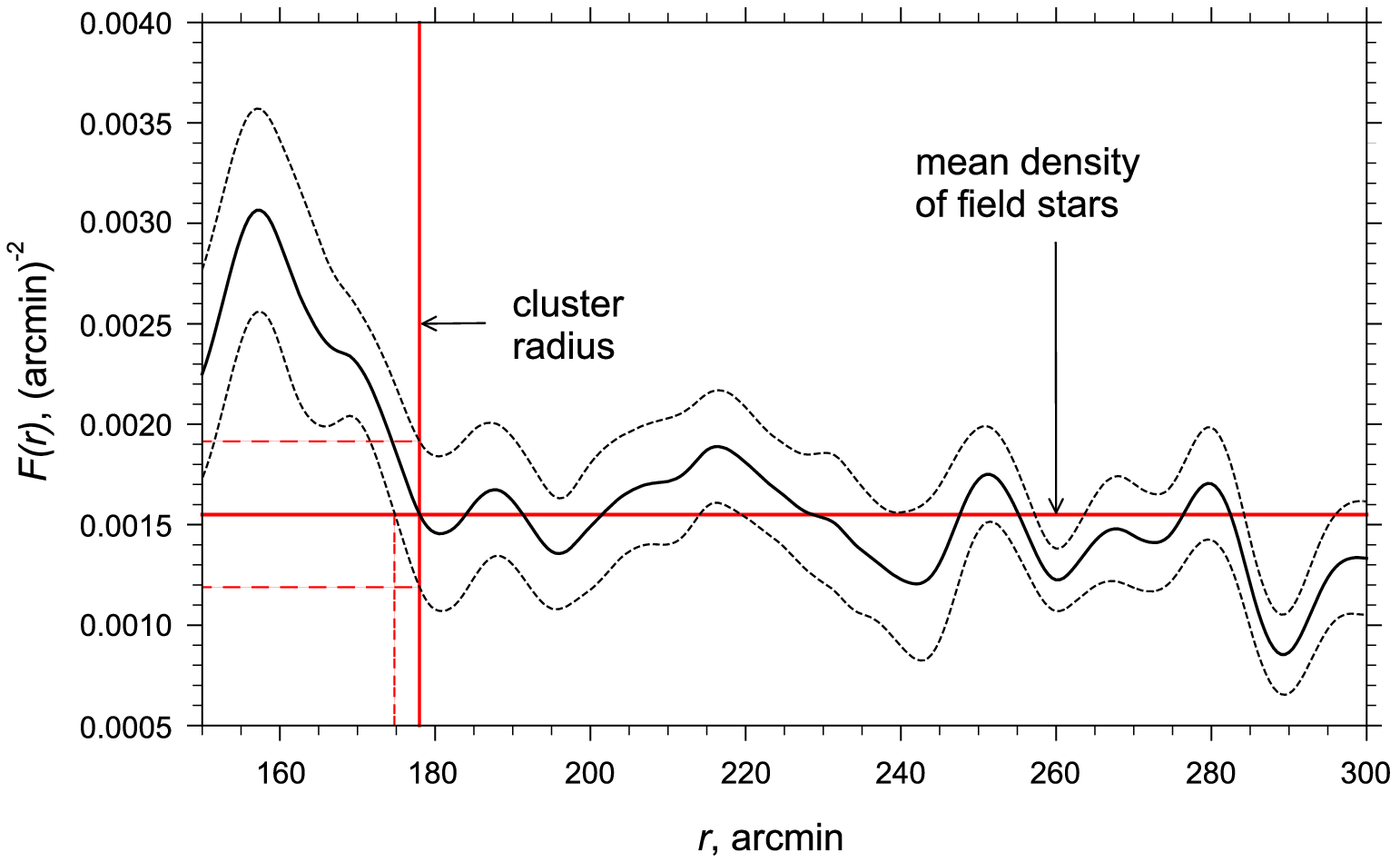}
\caption{Determination of NGC 3532 radius and the mean density of field stars (red solid lines) by the method \cite{Seleznev2016}. The black solid line is the density profile, the black dotted lines are  the boundaries of the confidence interval of $2\sigma$-width. The red dotted lines show evaluation of the errors of the radius and mean density. }
\label{radius1}
\end{figure}

As a first step, we determine the parallax and proper motion ranges of Gaia, which include (almost) all of the cluster stars and a relatively small number of background stars.
This work is performed in two stages.
First, we use a small region of 0.5$^\circ \times$0.5$^\circ$ around the cluster center $l_c=289.55795^\circ$ and $b_c=1.375277^\circ$ \cite{Dias+2021}.
To determine the ranges, we exploit the diagrams `parallax $\varpi$ -- magnitude G', `proper motion in right ascension $\mu_\alpha$ -- magnitude G', and `proper motion in declination $\mu_\delta$ -- magnitude G'.
We chose the range boundaries visually, so that probable cluster members would not fall outside the range.
That is, the range should include the cluster strip on the `parameter -- magnitude G' diagram with a reserve, taking into account the expansion of the strip when approaching $G=18$ mag.
At the same time, we control the width of the range of distances from the cluster center along the line of sight and the width of the range of velocities relative to the average cluster velocity, that correspond to the selected ranges of parallaxes and proper motions.
We take the distance value (477±2 pc) and the average values of the cluster proper motions ($\overline{\mu_\alpha}=-10.373\pm0.409$ mas/yr and $\overline{\mu_\delta}=5.193\pm0.430$ mas/yr) from the catalog of Dias et al. \cite{Dias+2021}.

As an example, Fig. \ref{plx-G_1} shows the `parallax $\varpi$ -- stellar magnitude G' diagram for the region around the cluster center.
The red lines in Fig. \ref{plx-G_1} show the selected range for the parallax of the stars of NGC 3532.

As a result, we obtained the following ranges for parallaxes and proper motions: 
\begin{equation}
\begin{array}{lll}
\varpi\in[1.262;2.862] \; \mbox{mas}  \\
\mu_\alpha\in[-13.338;-7.408]  \; \mbox{mas/yr} \\
\mu_\delta\in[2.228;8.158]  \; \mbox{mas/yr} .
\end{array}
\label{interval_1}
\end{equation}

These ranges correspond to a distance range of [349;792] pc and a velocity range of $\pm$6.7 km/s relative to the average velocity of the cluster.

In the next step, we use stars from the Gaia DR3 catalog from the 6.0$^\circ \times$6.0$^\circ$ region around the cluster center that satisfy the constraints obtained on the parallax and proper motions.
For these stars, we again plot the $\varpi$ -- G, $\mu_\alpha$ -- G, $\mu_\delta$ -- G diagrams to refine the range boundaries and cut off more background stars using the same approach as in the first step.
As an example, Fig. \ref{plx-G_2} shows the `parallax $\varpi$ -- magnitude G' diagram for stars from the 6.0$^\circ \times$6.0$^\circ$ region that satisfy the constraints obtained above.
The red lines show the refined parallax range within which the cluster stars are located.

Final ranges for parallaxes and proper motions:
\begin{equation}
\begin{array}{lll}
\varpi\in[1.725;2.399] \; \mbox{mas}  \\
\mu_\alpha\in[-11.646;-9.100]  \; \mbox{mas/yr} \\
\mu_\delta\in[4.000;6.386]  \; \mbox{mas/yr} .
\end{array}
\label{interval_2}
\end{equation}

These ranges correspond to a distance range of [417;580] pc and a velocity range of $\pm$2.9 km/s relative to the average cluster velocity in the right ascension direction and $\pm$2.7 km/s relative to the average cluster velocity in the declination direction.
Obviously, these ranges of distances and velocities are much wider than the possible deviations of the distances and velocities of probable open cluster members in the coordinate and the velocity space, respectively.
Thus, the probability that a cluster member star falls outside the ranges
(\ref{interval_2}) is very small.

For further study, we select stars from the Gaia DR3 catalog in the region 12.0$^\circ \times$12.0$^\circ$ around the cluster center, that satisfying the constraints (\ref{interval_2}).
We determined the cluster radius using the radial surface density profile, following the method of \cite{Seleznev2016} based on the kernel density estimator (KDE) \cite{Silverman1986,Merritt&Tremblay1994}.
Fig. \ref{profile} shows The radial surface density profile.
To plot it, we used a biquadratic kernel with a half-width of $h=10$ arcmin.
The optimal value of the kernel half-width was estimated using the method of \cite{Merritt&Tremblay1994,Seleznev2016}.
The dashed lines in Fig. \ref{profile} show the confidence interval of the $2\sigma$-width plotted using the smoothed bootstrap method \cite{Merritt&Tremblay1994,Seleznev2016}.
Fig. \ref{radius1} illustrates the determination of the average field star number density and the cluster radius using the method of \cite{Seleznev2016}.

According to our results, the radius of the cluster NGC 3532 is $R_c=178\pm3$ arcminutes, the average number density of field stars is $F_b=0.0016\pm0.0004$ stars per square arcminute.

To calculate the number of stars in the cluster $N_c$, we select stars in a circle of radius $R_c$ from the stars already selected in the region 12.0$^\circ \times$12.0$^\circ$ (N=2891 stars).
There were 2366 such stars in the circle.
The number of field stars $N_b$ in the circle of radius $R_c$ is $N_b=\pi R_c^2F_b$.
Subtracting the field stars $N_b=159$ from the stars in a circle with the cluster radius, we obtain the number of stars in the cluster $N_c$=2207 stars.
To estimate the error in the number of the cluster stars, we perform similar actions, using the maximum and minimum values of the cluster radius and the maximum and minimum values of the average field density within the error bounds of these quantities.
We use the maximum deviation from the average as the final error in the number of stars.
Finally, we obtain the number of stars in the cluster $N_c=2200\pm40$ stars.
Recall that this is the number of stars in the cluster that have 5- and 6-parameter solutions in the Gaia DR3 catalog.

The list of 2366 stars within a circle with radius $R_c$ is hosted in cloud storage at 
https:
//drive.google.com/file/d/1Z6-9DgLTJWtniDBCOTgbp76ZdXcEgWnC/view?usp=sharing.
Using the uniform field method \cite{Danilov2020}, we can estimate the mean probability of these stars to be the cluster member as 93\%.

Note that the circle with radius $R_c$ includes almost all the stars from the Hunt\&Reffert \cite{H&R2023} sample (only three stars from their sample are located at a greater distance from the cluster center).
The number of stars $N_c=2200\pm40$ is in a good agreement with the total number of stars in NGC 3532 from their sample with $G<18$ mag (2158) and is greater than the number of stars in this cluster with a membership probability of more than 50\% at the same magnitude limit (1505).
At the same time, our sample contains 2120 stars of the Hunt\&Reffert sample \cite{H&R2023}, which is 98.2\% of their sample.
Thus, the very simple method of selecting stars for star counts that we used gave a result that is in a good agreement with the result of the more complicated method \cite{H&R2023}.
The average value of the membership probability in our sample (2366 stars) obtained by the uniform field method is quite high (93\%).
There are only 76 stars with such probability in Hunt\&Reffert's \cite{H&R2023} sample, but they used a completely different method to estimate the probability.

\section{SEARCH FOR POSSIBLE NGC 3532 MEMBERS WITH POOR ASTROMETRIC SOLUTIONS}

To look for possible members of the NGC 3532 cluster that were not included in the sample obtained above, we selected stars in Gaia DR3 from the region 10.0$^\circ \times$10.0$^\circ$ around the cluster center that have only 2-parameter solutions, stars with the parameter RUWE>1.4, and stars with relative parallax errors greater than 0.2.

\begin{figure}[ht]
\includegraphics[width=0.5\textwidth]{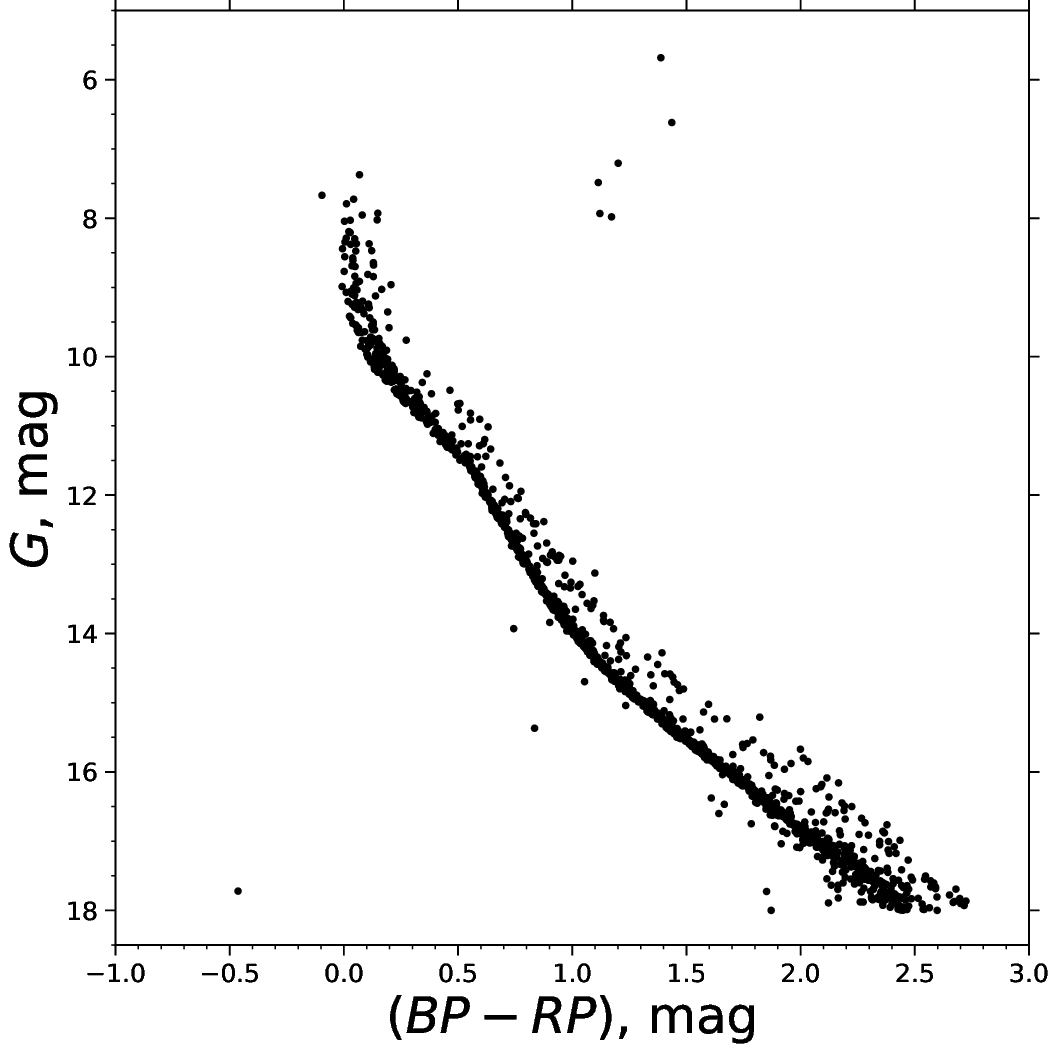}
\caption{CMD for stars from the Hunt\&Reffert \cite{H&R2023} sample with a membership probability greater than 50\%. }
\label{CMD}
\end{figure}

\begin{figure}[ht]
\includegraphics[width=0.5\textwidth]{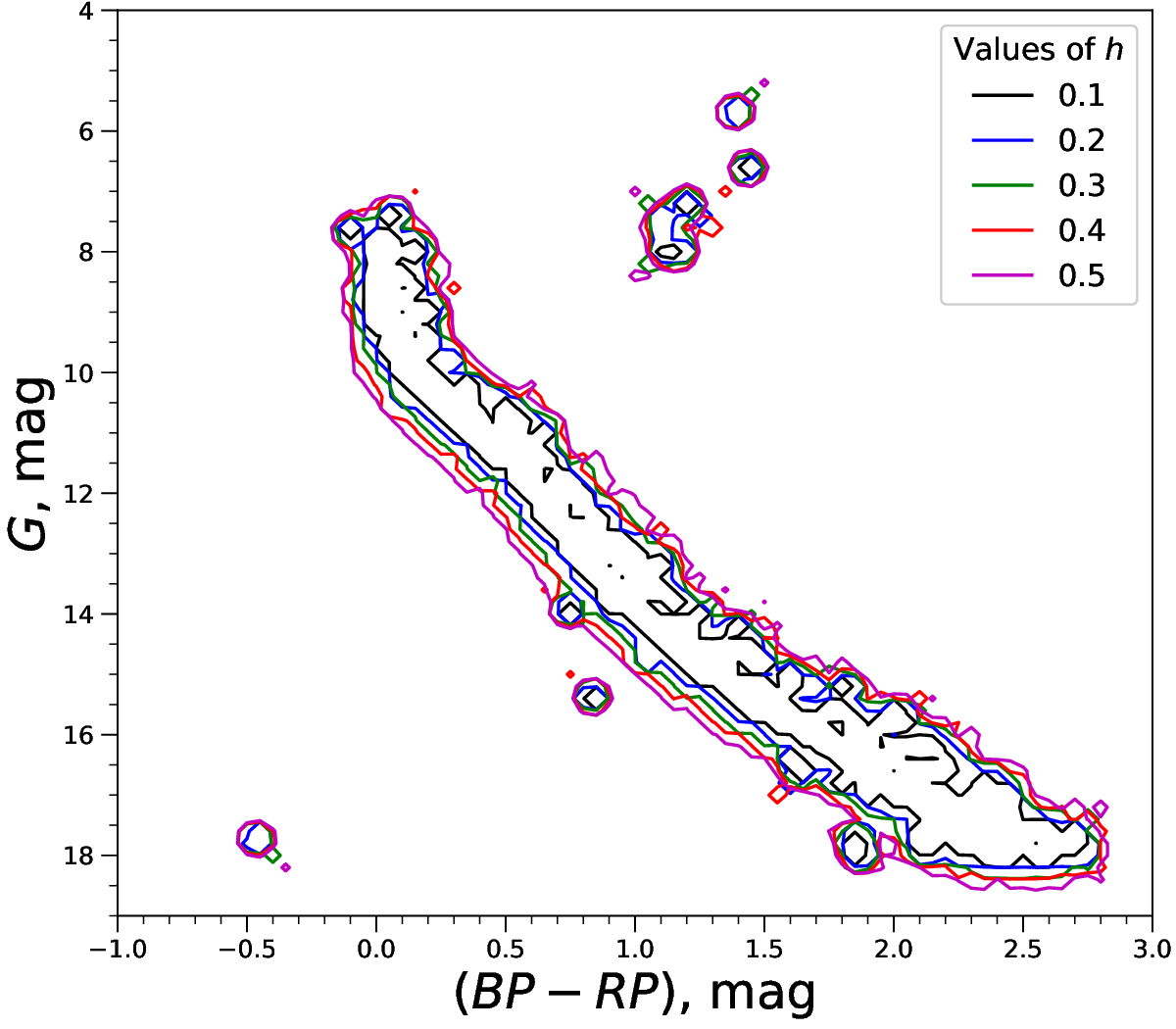}
\caption{Hess diagram for different values of the kernel half-width $h$. The lines of different colors show the isoline of the probability density of 0.01 for different values of $h$ shown in the legend. }
\label{Hess}
\end{figure}

\begin{figure}[ht]
\includegraphics[width=0.6\textwidth]{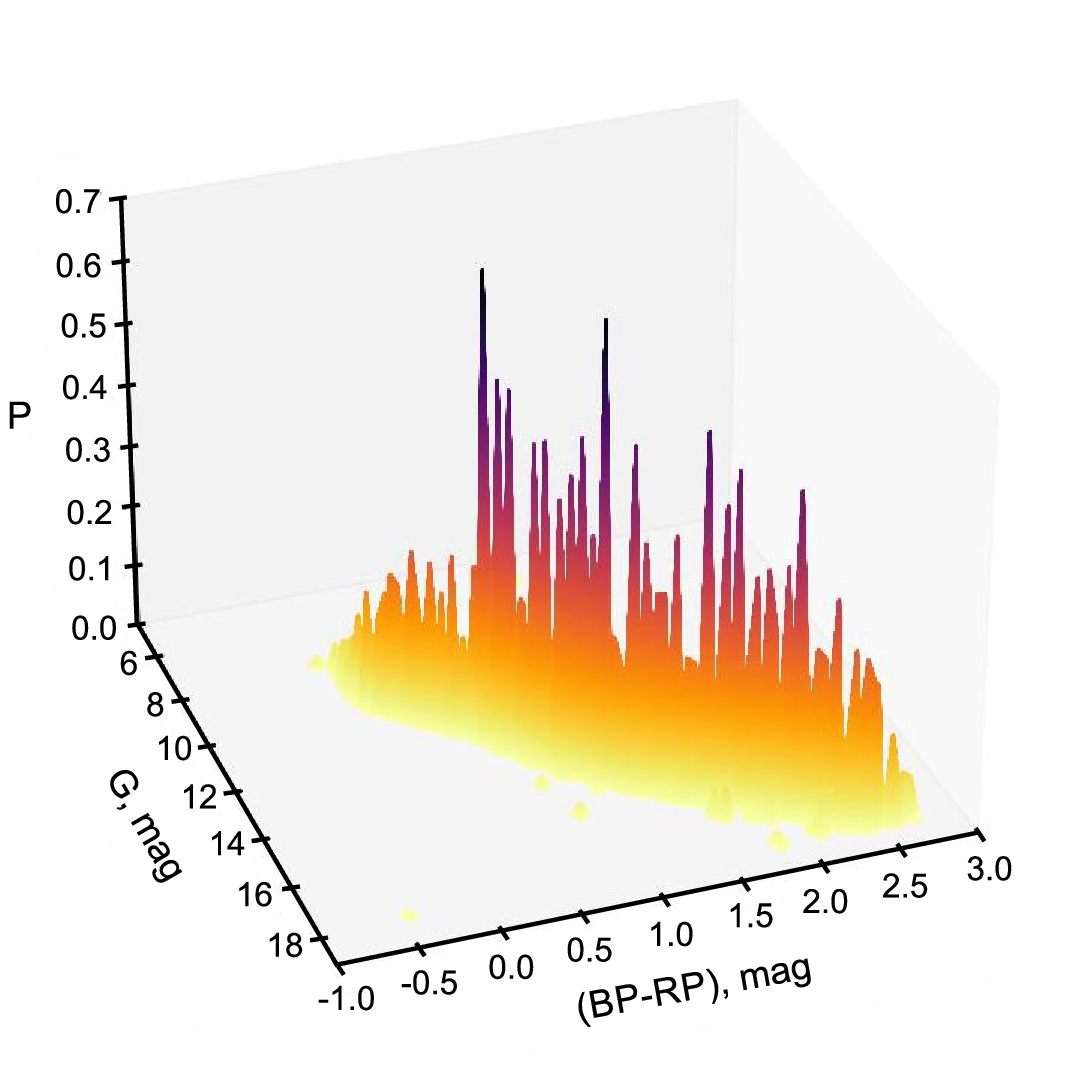}
\caption{Three-dimensional image of the Hess diagram for the kernel half-width $h=0.1$ mag. Darker areas indicate higher probability. }
\label{Hess1}
\end{figure}

\begin{figure}[ht]
\includegraphics[width=0.6\textwidth]{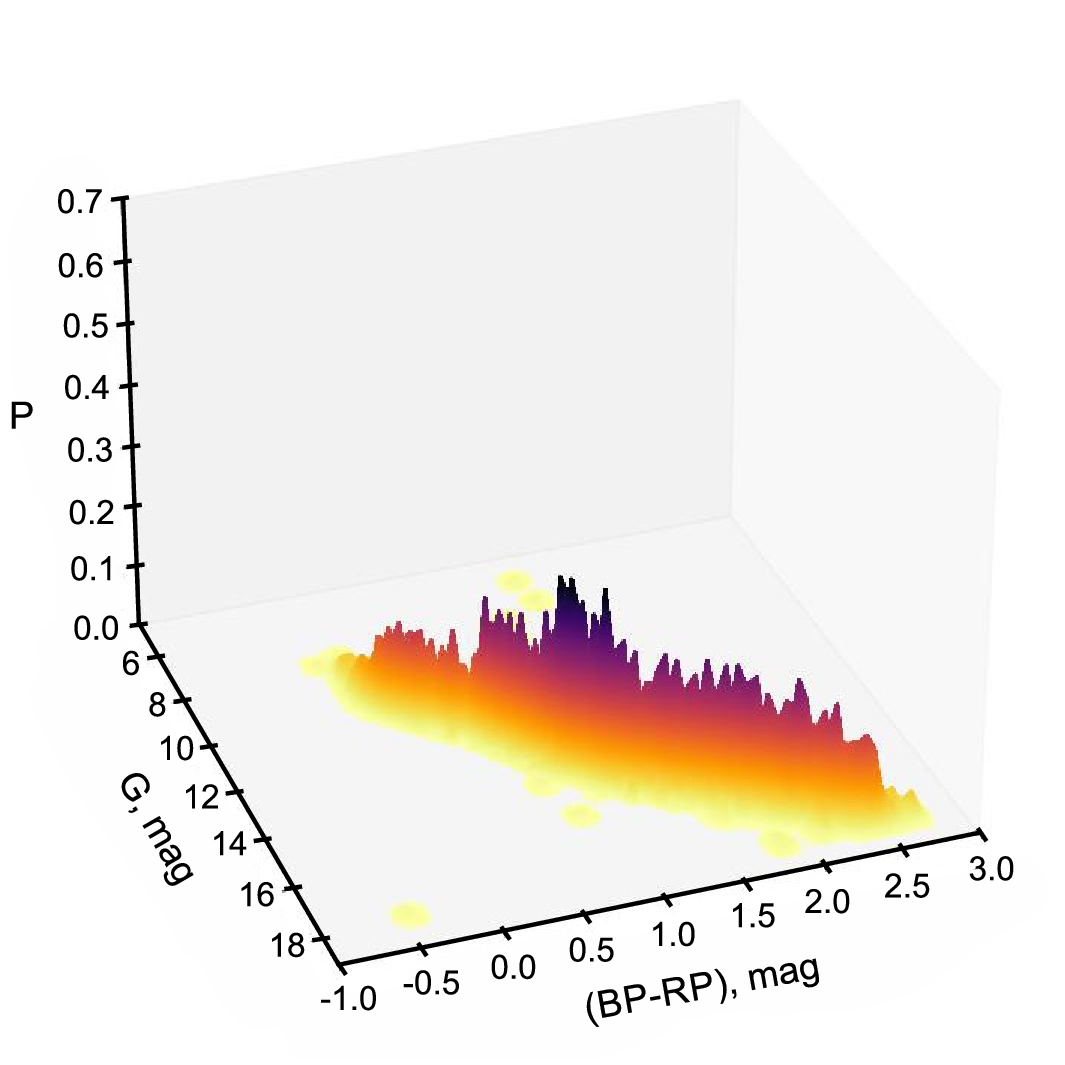}
\caption{The same as in Fig. \ref{Hess1}, but for $h=0.3$ mag. }
\label{Hess2}
\end{figure}

\begin{figure}[ht]
\includegraphics[width=0.6\textwidth]{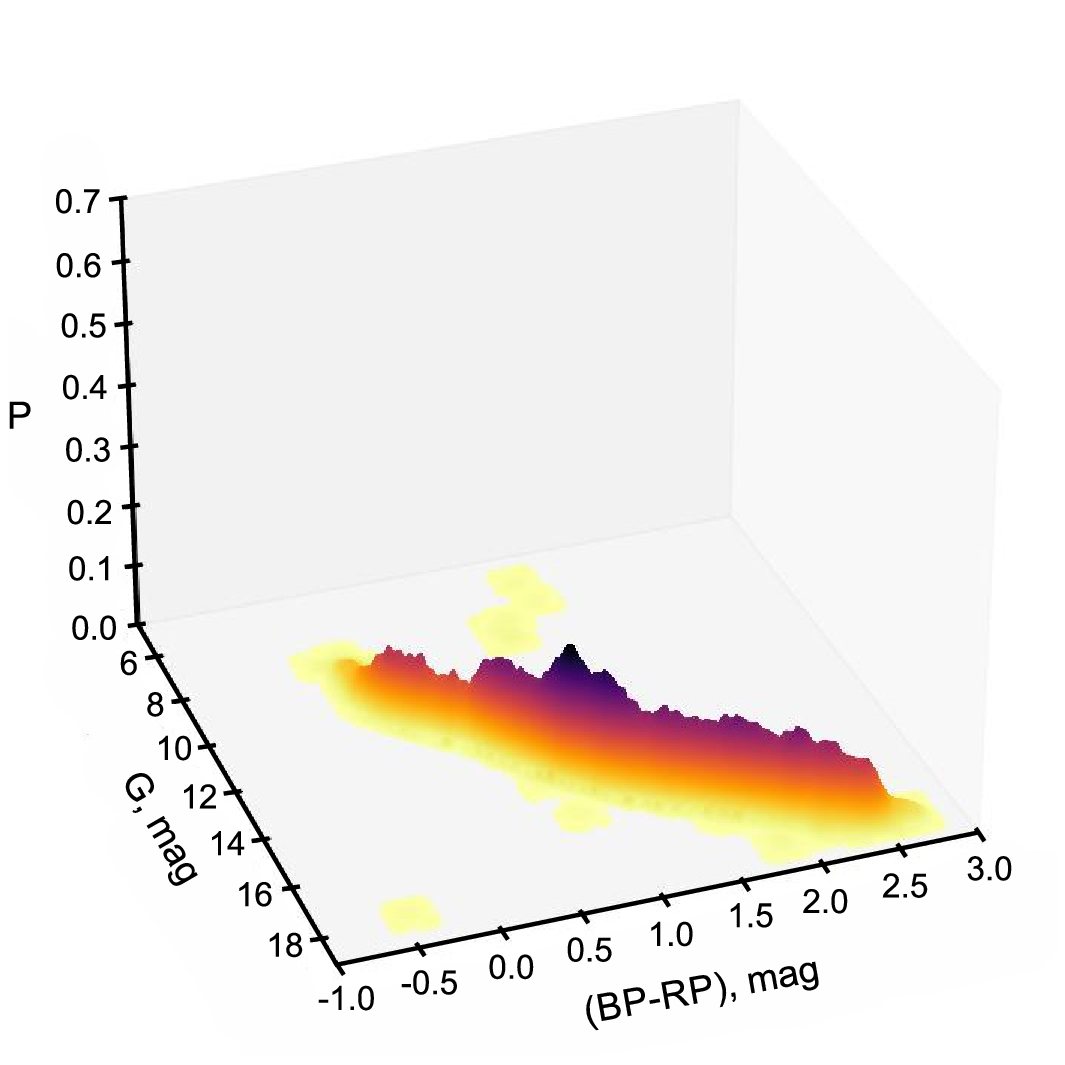}
\caption{The same as in Fig. \ref{Hess1}, but for $h=0.5$ mag. }
\label{Hess3}
\end{figure}

\begin{figure}[ht]
\includegraphics[width=0.6\textwidth]{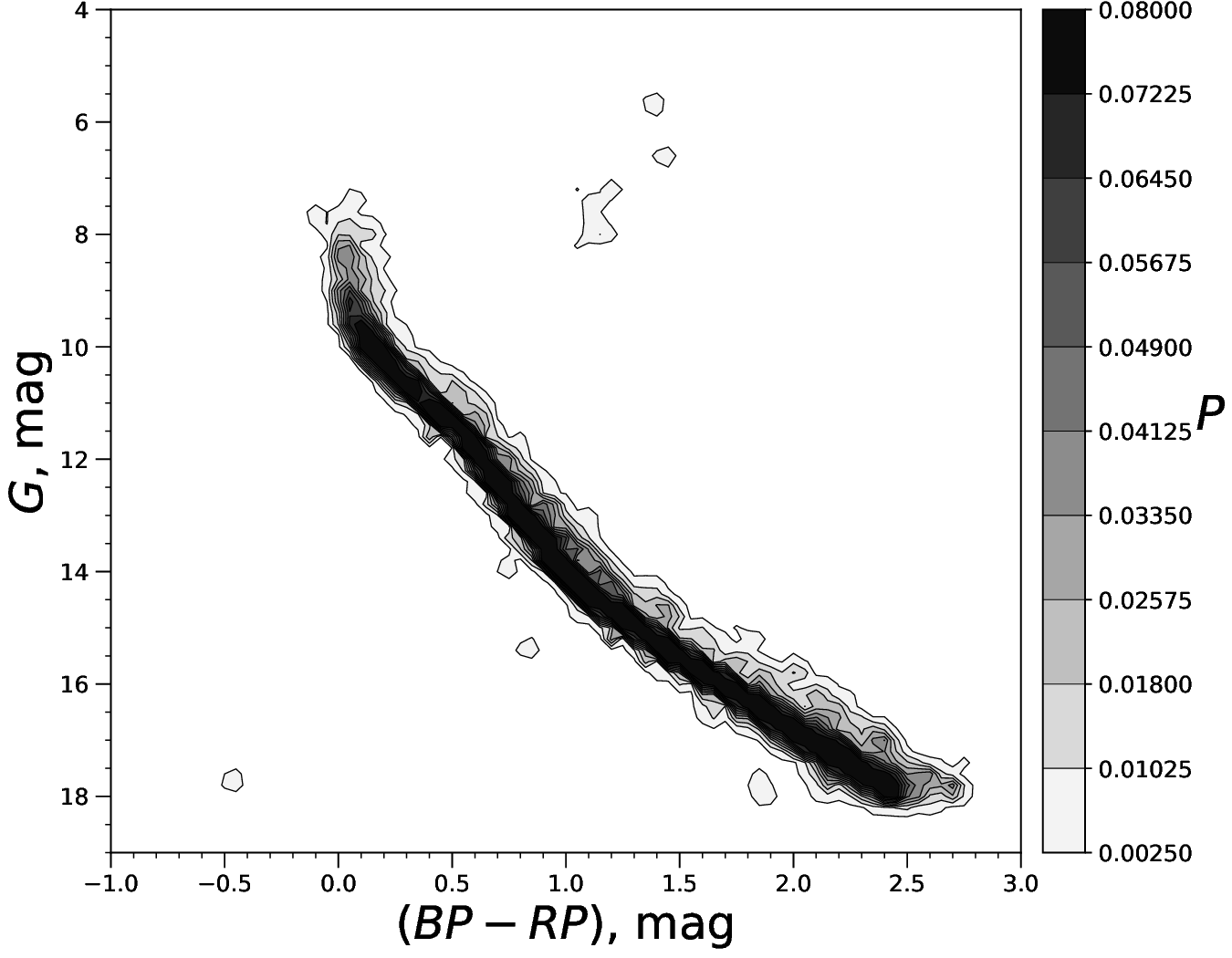}
\caption{Hess diagram for NGC 3532. Probability values are shown on the scale on the right. }
\label{Hess_final}
\end{figure}

As a criterion for possible membership of a star in the cluster, we use the star's inclusion in the region occupied by the cluster on the CMD.
In this case, we can even estimate the probability of belonging to this region using the Hess diagram --- the probability density distribution on the CMD.
To plot the Hess diagram, we took stars from the Hunt\&Reffert \cite{H&R2023} sample with a membership probability greater than 50\%.
The CMD for these stars is shown in Fig. \ref{CMD}.

To plot the Hess diagram we use the two-dimensional KDE method with a biquadratic kernel \cite{Silverman1986}:
\begin{equation}
\label{kernel}
    K(x)= \left\{
            \begin{array}{ll}
\frac{\displaystyle 3}{\displaystyle \pi h^2}\left(1-\frac{\displaystyle x^2+y^2}{\displaystyle h^2}\right)^2 & ,\; x^2+y^2\leq h^2 \\
0                & ,\; x^2+y^2 > h^2 . \\
            \end{array}
    \right.
\end{equation}

\noindent Density values are calculated at the nodes of a rectangular grid.
To go from density values to probability values, it is necessary to divide all density values at the nodes by the total number of points used to plot the Hess diagram.

When using the KDE method, an important point is the choice of the optimal kernel half-width $h$.
To estimate the optimal half-width, we plotted a Hess diagram for different values of the half-width.
Fig. \ref{Hess} shows the isoline of the probability density of 0.01 for different values of $h$ by lines of different colors.
It is easy to see that the size of the region in the Hess diagram where possible cluster members fall does not depend very much on $h$.

The spread of probability values within this region depends much more on $h$.
This is clearly seen in Fig. \ref{Hess1}-\ref{Hess3}, which show three-dimensional images of the Hess diagram for three values of the kernel half-width.
For small values of the half-width, the probability values reach large values, but the spread of probability values also is large.
For large values of the half-width, the spread becomes smaller, but the maximum probability values also are small.
We list the maximum probability values depending on the kernel half-width in Table \ref{h_prob}.
Finally, we settled on the kernel half-width value for Hess diagram as $h=0.3$.

Before using Eq. (\ref{kernel}), we should change the scale of a color index (BP-RP) to align the scales along the axes.
Otherwise, the kernel would not be spherical.
We used the formula
\begin{equation}
\label{kernel_norm}
    (BP-RP)'=((BP-RP)+1)\cdot4 \; .
\end{equation}

\noindent We added the unit for convenience, to work only with positive numbers.
When plotting the Hess diagram and for further research, we performed the inverse transformation.
Fig. \ref{Hess_final} shows the final version of the Hess diagram.
The scale on the right side of the figure shows the range of probability values.
We chose a shortened probability range for Fig. \ref{Hess_final} in order to show better the areas of the diagram with smaller probability values.

To search for stars with poor astrometric solutions that might be members of the NGC 3532 cluster, we selected stars from Gaia DR3 in the region 10.0$^\circ \times$10.0$^\circ$ around the cluster center that satisfy the following constraints:
\begin{itemize}
    \item $G<18$ mag (the star must have photometry data in three bands);
    \item star is not in the sample \cite{H&R2023}.
\end{itemize}
We selected stars with 2-parameter solutions or stars with 5- or 6-parameter solutions, but having the parameter RUWE>1.4, and/or the relative parallax error $\varpi/\delta_\varpi>0.2$.
There were about 1.5 million of such stars in the study area.

To determine the probability of a star to fall into the cluster region on the Hess diagram, we performed the linear interpolation over the grid nodes at the corners of the cell into which the star fell.

We took 0.00705 as the lower probability value.
It corresponds to the maximum value of the \ref{kernel} function.
This cuts off stars that fall into areas on the Hess diagram created by only one star.
As a result, we obtained a sample of 30,587 stars that fall into the area occupied by the cluster on the Hess diagram with a probability greater than 0.00705.
Among these stars 14,801 stars are inside the circle of the radius equal to the cluster radius obtained by stars with poor astrometric solutions (see below).
The catalog of these 14,801 stars is located in the cloud storage at https://drive.google.com/file/d/1Z6-9DgLTJWtniDBCOTgbp76ZdXcEgWnC/view?usp=sharing.

It is obvious that this list contains a large number of field stars.
In order to estimate the number of possible cluster members, we applied the same method as in Section II to the list of stars with poor astrometric solutions.

\begin{figure}[ht]
\includegraphics[width=0.7\textwidth]{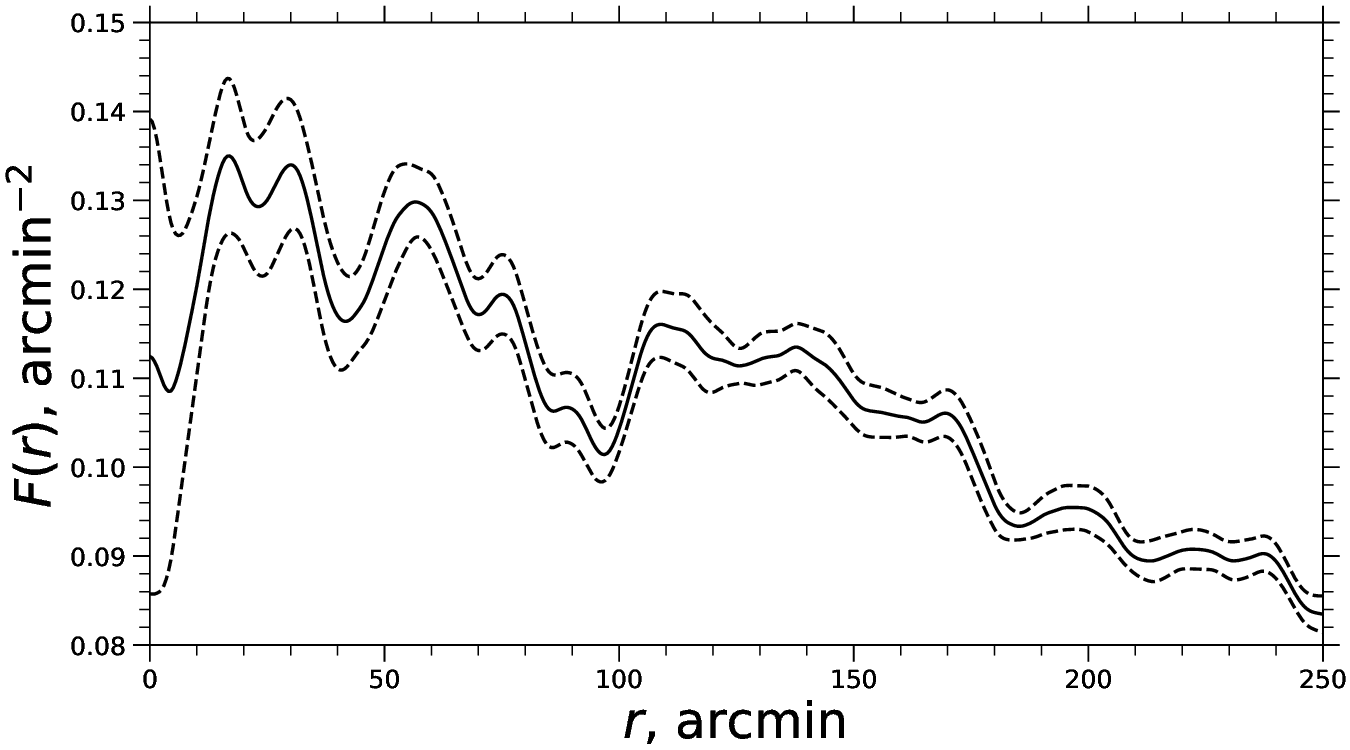}
\caption{Surface density profile for stars with poor astrometric solutions, selected using the Hess diagram. The notations are the same as in Fig. \ref{profile}. }
\label{profile_bad}
\end{figure}

\begin{figure}[ht]
\includegraphics[width=0.7\textwidth]{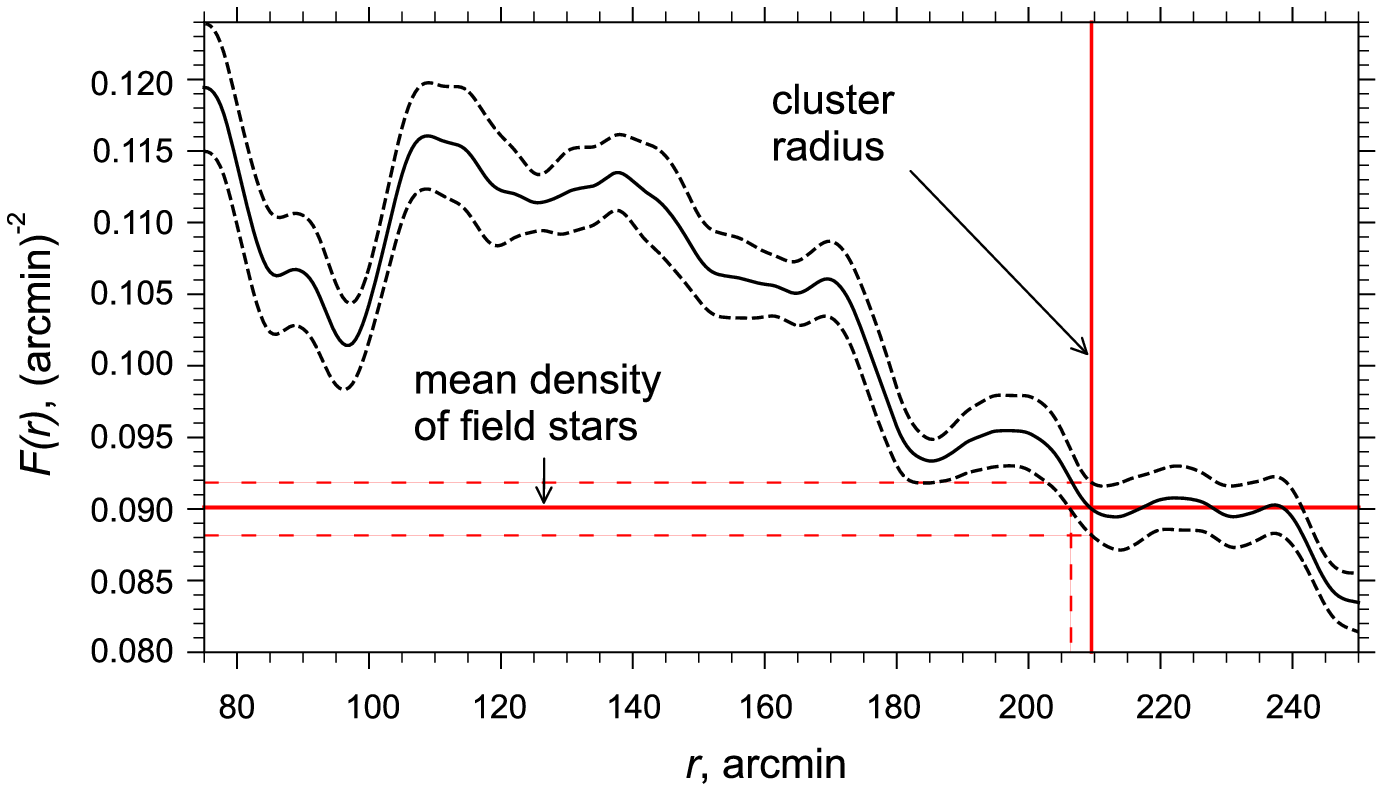}
\caption{Determination of NGC 3532 radius and the mean density of field stars by the method of \cite{Seleznev2016} by stars with poor astrometric solutions Gaia DR3. Designations are the same as in Fig. \ref{radius1}. }
\label{radius2}
\end{figure}

\begin{figure}[ht]
\includegraphics[width=0.6\textwidth]{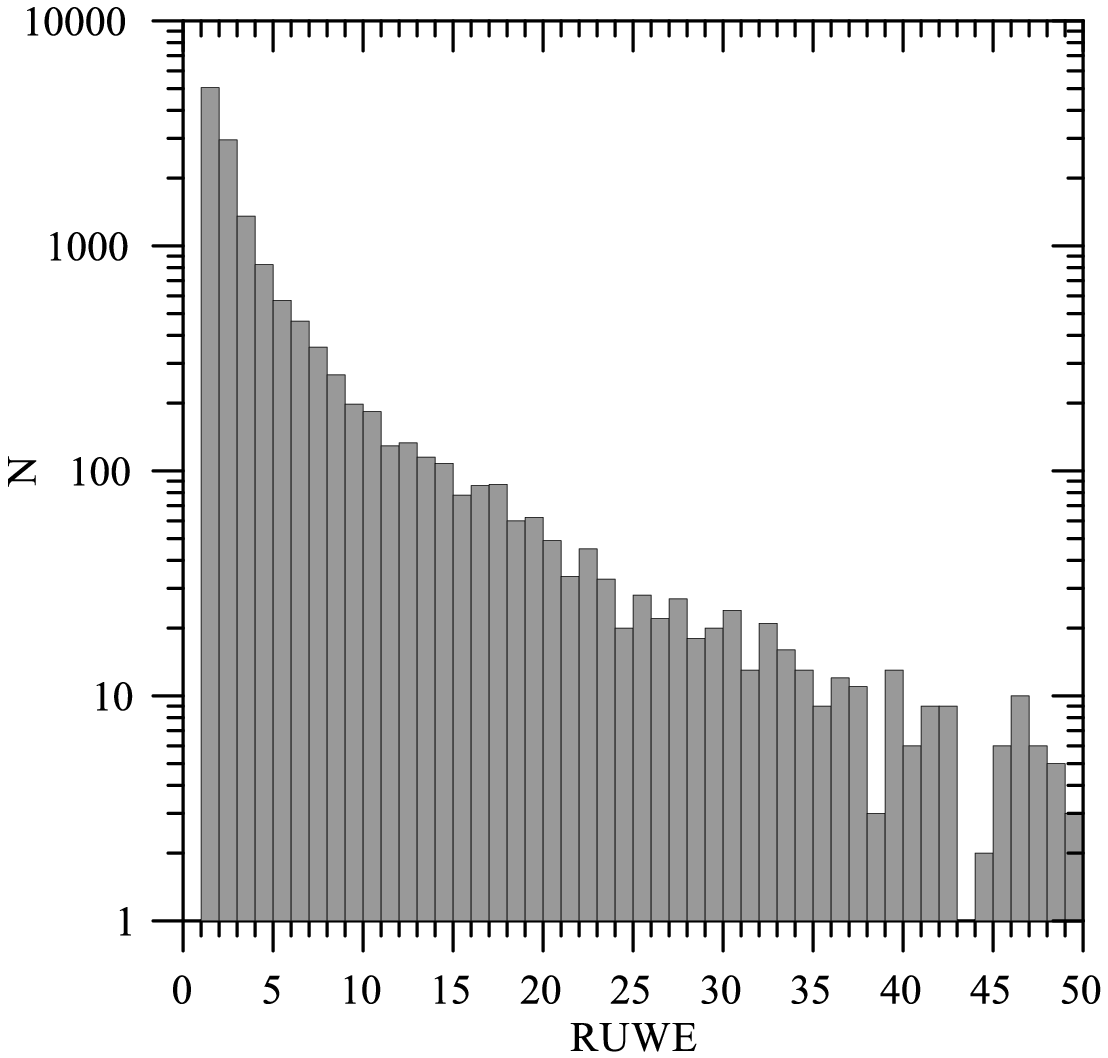}
\caption{Distribution of stars with poor astrometric solutions by parameter RUWE. }
\label{RUWE_distr}
\end{figure}

\begin{figure}[ht]
\includegraphics[width=0.7\textwidth]{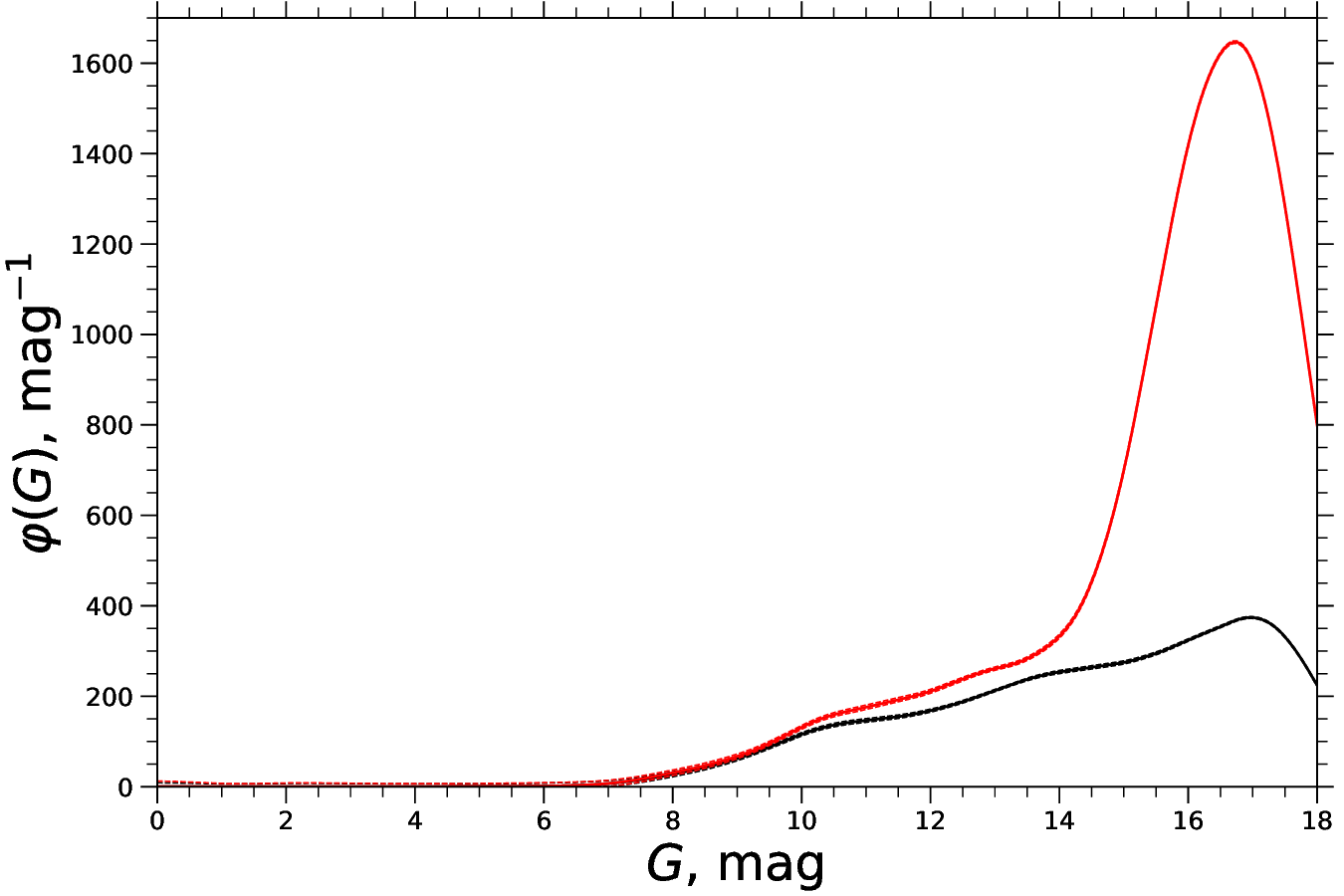}
\caption{Luminosity functions for the cluster NGC 3532. The solid lines --- the luminosity functions, the dotted lines --- $2\sigma$ confidence intervals. Black lines --- for stars with 5- and 6-parameter solutions ($\overline\sigma=4.7$ (mag)$^{-1}$), red lines --- with an addition of stars with poor astrometric solutions ($\overline\sigma=5.6$ (mag)$^{-1}$).} 
\label{LF}
\end{figure}

\begin{table}
\caption{Dependence of the maximum probability value on the Hess diagram on the kernel half-width}
\label{h_prob}
\begin{tabular}{ c|c }
\hline
Kernel halfwidth $h$ & Maximum probability value \\
\hline

0.1                    &  0.750                           \\
0.2                    &  0.433                           \\
0.3                    &  0.320                           \\
0.4                    &  0.265                           \\
0.5                    &  0.212                           \\

\hline
\end{tabular}
\end{table}

To do this, we plotted a radial profile of the surface density for the stars in this list, using KDE with a biquadratic kernel and a kernel half-width of 10 arcmin.
This profile is shown in Fig. \ref{profile_bad}.
It is obvious that there is a concentration of stars towards the center of the cluster, although with some dip in the very center.

Fig. \ref{radius2} shows the determination of the cluster radius and the average density of field stars based on stars with poor astrometric solutions.
We took the interval of distances from the center between 210 and 240 arcminutes from the center as the field region.
The decrease in the surface density at $r>240$ arcminutes is most likely due to the large-scale field density fluctuations around the cluster.
In Section II, the sample for plotting the density profile contains fewer field stars, and large-scale fluctuations of the field density do not have a noticeable effect.
In the case of stars with poor astrometric solutions, the estimated cluster radius was $R_{c2}=210\pm3$ arcminutes and the estimated average number density of field stars was $F_{b2}=0.090\pm0.002$ stars per square arcminute.

We estimated the number of possible members of the NGC 3532 cluster with poor astrometric solutions using the same method as in Section II.
It turned out to be $N_{c2}=2150\pm230$.
Thus, the number of stars with poor astrometric solutions in the NGC 3532 cluster is very close to the number of cluster members with `good' solutions $N_c=2200\pm40$.
Now we can estimate the membership probability of stars from the list obtained with the Hess diagram.
This probability is approximately 14.5\%.
For this, we used the uniform field method \cite{Danilov2020}, as in Section II.
It is very important that a large number of unresolved binary and multiple systems may be among the stars with poor astrometric solutions of Gaia DR3, since this circumstance is one of the main reasons to obtain a poor solution.

Among the stars with poor astrometric solutions (14801), 1202 stars have only a two-parameter solution, and 13599 stars have a large relative parallax error and a large RUWE value.
The distribution of these stars by RUWE is shown in Fig. \ref{RUWE_distr}.

How will the luminosity function for the NGC 3532 cluster change when we add the stars with poor astrometric solutions?
In Fig. \ref{LF}, the black solid line is the luminosity function for stars with 5- and 6-parameter solutions, and the red solid line is the luminosity function with an addition of the stars with poor astrometric solutions.
The dashed lines show confidence intervals of $2\sigma$-width.
We plotted the luminosity functions by the KDE method using a quartic (biquadratic) kernel \cite{Silverman1986,Merritt&Tremblay1994,KDE_OSC} with a half-width of 1.5 stellar magnitudes.
This explains the dip in the luminosity functions when they approach G = 18 magnitudes since we consider only stars up to this magnitude (see above).
To construct the luminosity function, we used the method proposed in \cite{LFmethod} and used, in particular, in \cite{Yeh2019,Danilov2020}.
We used a ring with an area equal to the cluster area and an inner radius equal to the cluster radius as a comparison region.

The luminosity functions differ greatly in the region of $G > 14$ magnitudes.
The luminosity function with addition of stars with poor astrometric solutions contains many more stars in the range $G\in[14,18]$ magnitudes.
The reason may be the fact that the errors of the astrometric parameters in Gaia DR3 increase noticeably at $G > 14$ magnitudes.
In any case, the fraction of binary and multiple stars among the possible members of the NGC 3532 cluster with poor astrometric solutions is noticeably greater than among the cluster stars with 5- and 6-parameter solutions.
This is confirmed by the distribution of stars with poor astrometric solutions by the RUWE parameter (Fig. \ref{RUWE_distr}), a large value of which is an indicator of a shift in the photometric center of a star due to the presence of a companion \cite{RUWE}.

\section{CONCLUSIONS}

In this paper, we performed star counts in a wide vicinity of the open star cluster NGC 3532 to determine how many stars with poor astrometric solutions of Gaia DR3 might be members of this cluster.

First, we performed star counts for stars with 5- and 6-parameter solutions and $G<18$ stellar magnitude, setting fairly strict constraints on the parallaxes and proper motions of the stars.
As a result, we estimated the cluster radius $R_c=178\pm3$ arcminutes.
Within a circle of such a radius, a sample of 2366 stars was obtained, of which $N_c=2200\pm40$ stars are members of the cluster NGC 3532.

To find possible cluster members with poor astrometric solutions, we used the criterion of a star falling to the region occupied by the cluster on the CMD.
For this purpose, we plotted a Hess diagram for the cluster NGC 3532, using the \cite{H&R2023} sample.

Then, we selected stars within a 10.0$^\circ \times$10.0$^\circ$ region around the cluster center with the same magnitude constraint that either have only a 2-parameter solution, or 5- and 6-parameter solutions, but have the parameter RUWE>1.4, and/or a large relative parallax error $\varpi/\delta_\varpi>0.2$.
Among these stars, we selected stars that fall within the region occupied by the cluster on the Hess diagram, and estimated the probability of their belonging to this region.

It turned out that the stars with a probability of more than 0.705\% have a noticeable concentration towards the center of the cluster.
This gave us the opportunity to use the method \cite{Seleznev2016} to determine the cluster radius and number of cluster stars with poor astrometric solutions.
In this case, we obtained the radius value $R_{c2}=210\pm3$ arcminutes, slightly exceeding the radius value obtained from `good' stars, and the number of cluster stars $N_{c2}=2150\pm230$, coinciding within the error with the number of `good' stars.

The luminosity function with addition of stars with poor astrometric solutions differs greatly from the luminosity function plotted only for stars with 5- and 6-parameter solutions in the range of $G\in[14,18]$ stellar magnitudes.
Stars with poor astrometric solutions are generally the relatively faint stars with $G > 14$ stellar magnitudes.

Thus, it can be concluded that with the traditional method of selecting probable cluster members, about half of the possible members are lost due to the fact that these stars have poor astrometric solutions of Gaia DR3.
We present a list of stars with poor astrometric solutions that can be members of the NGC 3532 cluster with a probability of 14.5\%.
It is very important that among these stars there can be a large number of unresolved binary and multiple systems, since this circumstance is one of the main reasons to obtain a poor solution.
This is confirmed by the distribution of stars with poor astrometric solutions by the RUWE parameter, large values of which are considered as a highly probable sign of star binarity \cite{RUWE}.

\section*{FUNDING}
This work was supported by the Ministry of science and higher education of the Russian Federation by an agreement FEUZ-2023-0019.

\begin{acknowledgments}
This work has made use of data from the European Space Agency (ESA) mission {\it Gaia} (\url{https://www.cosmos.esa.int/gaia}), processed by the {\it Gaia} Data Processing and Analysis Consortium (DPAC, \url{https://www.cosmos.esa.int/web/gaia/dpac/consortium}). 
Funding for the DPAC has been provided by national institutions, in particular the institutions participating in the {\it Gaia} Multilateral Agreement.

\end{acknowledgments}

\section*{CONFLICT OF INTEREST}

The authors declare no conflict of interest.

\end{document}